\documentclass[a4paper,12pt]{iopart}
\usepackage{graphicx}
\usepackage{latexsym}
\usepackage{amsfonts,amsbsy}

\def\Gammabol{{\stackrel{\circ}{\Gamma}}{}}

\def\be{\begin{equation}}
\def\ee{\end{equation}}
\def\ba{\begin{eqnarray}}
\def\ea{\end{eqnarray}}
\begin{document}

\jl{6}

\title{Gravitation: global formulation and quantum effects}

\author{R Aldrovandi, J G Pereira and K H Vu}
\vskip 0.3cm
\address{Instituto de F\'{\i}sica Te\'orica, 
Universidade Estadual Paulista,
Rua Pamplona 145,
01405-900\, S\~ao Paulo SP, Brazil}

\begin{abstract}
A nonintegrable phase-factor global approach to gravitation is developed by
using the similarity of teleparallel gravity with electromagnetism. The phase
shifts of both the COW and the gravitational Aharonov-Bohm effects are obtained.
It is then shown, by considering a simple slit experiment, that in the classical
limit the global approach yields the same result as the gravitational Lorentz
force equation of teleparallel gravity. It represents, therefore, the {\it
quantum mechanical} version of the {\it classical} description provided by the
gravitational Lorentz force equation. As teleparallel gravity can be formulated
independently of the equivalence principle, it will consequently require no
generalization of this principle at the quantum level.
\end{abstract}

\pacs{04.20.-q; 11.15.-q}
\maketitle

%%%%%%%%%%%%%%%%%%%%%%
\section{Introduction}

Gauge potentials (or connections), which in classical theories are simply
convenient mathematical tools from which the real physical fields can be
obtained, assume a fundamental role in quantum theories. The conceptual changes
occurring in the passage from classical to quantum theory include the larger
importance acquired by the canonical formalism (written in terms of the
potentials) and the minor role left to the idea of force in comparison with the
concepts of momentum and energy, which become crucial. Instead of trajectories,
quantum mechanics considers probability amplitudes  obtained from 
wavefunctions, whose  wavelengths are associated to momentum and energy. Instead
of forces,  quantum mechanics deals with the way these wavelengths  are changed
by the interactions. The significance of these conceptual issues is clearly
manifest in the Aharonov-Bohm effect~\cite{ab}, a quantum interference
phenomenon taking place in a (non simply-connected) space region in which the
electromagnetic field---but not the gauge potential---vanishes. It cannot be
described by the classical Lorentz force equation, and imposes the use of a
quantum approach, as for example the global formulation of electromagnetism 
based on the notion of nonintegrable phase-factors~\cite{wy}.

On the other hand, the teleparallel equivalent of general relativity
\cite{tegra}, or teleparallel gravity for short \cite{obs,op}, can be understood
as a gauge theory for the translation group~\cite{PR}. The fundamental field
describing gravitation according to this theory is the translational gauge
potential\footnote{We use the Greek alphabet $\mu, \nu, \rho, \dots = 0, 1, 2, 3$
to denote spacetime indices, and the Latin alphabet $a, b, c, \dots = 0, 1, 2,
3$ to denote anholonomic indices related to the tangent Minkowski spaces, whose
metric is chosen to be $\eta_{a b} = {\rm diag} (+1, -1, -1, -1)$.} $B^a{}_\mu$,
a 1-form assuming values in the Lie algebra of the translation group \cite{wep}.
By using the analogy with the electromagnetic gauge potential $A_\mu$, which is
also an Abelian gauge potential, it becomes possible to define a gravitational
nonintegrable phase-factor. In other words, it becomes possible to construct a
teleparallel global formulation for gravitation. This will be the main purpose
of this paper. We begin by introducing, in section 2, the strictly necessary
ingredients of teleparallel gravity. Then, in section 3, we study the Newtonian
limit of the equation of motion of a spinless particle of mass $m$. In
teleparallel gravity, it is given by a force equation \cite{p1} similar to the
Lorentz force equation of electrodynamics which, when written in terms of the
Christoffel connection, coincides with the geodesic equation of general
relativity. In section 4, in analogy with electromagnetism, we develop, in terms
of a nonintegrable phase-factor, a teleparallel global formulation for
gravitation. This formulation represents the {\em quantum mechanical} law that
replaces the {\em classical} gravitational Lorentz force equation of
teleparallel gravity. As a first application, we use this global formulation to
deduce, in section 5, the well known results of the Colella-Overhauser-Werner
(COW) experience \cite{cow}. Then, as a second application, in section 6, we use
the same formulation to obtain the phase shift related to the gravitational
analog of the Aharonov-Bohm effect. In section 7 it is shown that, in the
classical limit, the quantum  phase-factor approach yields the same results as
the classical approach of the gravitational Lorentz force equation, which shows
the consistency of the global formulation of gravitation. Finally, in section 8,
we sum up the results obtained. 

%%%%%%%%%%%%%%%%%%%%%%%%%%%%%%
\section{Teleparallel Gravity}

As already mentioned, the fundamental field describing gravitation in
teleparallel gravity is the translational gauge potential $B^a{}_\mu$, a 1-form
assuming values in the Lie algebra of the translation group,
\be
B_\mu = B^a{}_\mu \, P_a,
\ee
with $P_a = \partial_a$ the generators of infinitesimal translations.
Analogously to the electromag\-ne\-tic case, the gravitational field strength is
given by
\be
T^a{}_{\mu \nu} = \partial_\mu B^a{}_\nu - \partial_\nu B^a{}_\mu.
\label{gfs}
\ee
In teleparallel gravity, the gauge potential $B^a{}_\mu$ appears as the
nontrivial part of the tetrad field $h^{a}{}_{\mu}$:
\be
h^a{}_\mu = \partial_\mu x^a + B^a{}_\mu.
\label{tetrada}
\ee
Notice that, whereas the tangent space indices are raised and lowered with the
Minkowski metric $\eta_{a b}$, the spacetime indices are raised and lowered with
the spacetime metric
\be
g_{\mu \nu} = \eta_{a b} \; h^a{}_\mu \; h^b{}_\nu.
\label{gmn}
\ee
Under a local translation of the tangent space coordinates $\delta x^a =
\epsilon^a(x)$, the gauge potential transforms according to
\be
B^{\prime a}{}_\mu = B^a{}_\mu - \partial_\mu \epsilon^a.
\label{btrans}
\ee
It is then an easy task to verify that both $T^a{}_{\mu \nu}$ and $h^a{}_\mu$
are invariant under this transformation.

Now, the tetrad $h^a{}_\mu$ gives rise to the so called Weit\-zen\-b\"ock
connection
\begin{equation}
\Gamma^{\rho}{}_{\mu\nu} = h_{a}{}^{\rho}\partial_{\nu}h^{a}{}_{\mu},
\label{carco}
\end{equation}
which introduces distant parallelism on the four-dimensional spacetime
manifold. It is a connection that presents torsion, but no curvature. Its
torsion
\begin{equation}
T^{\rho}{}_{\mu\nu} = \Gamma^{\rho}{}_{\nu\mu} - 
\Gamma^{\rho}{}_{\mu\nu},
\label{tor}
\end{equation}
as can then be easily checked, is related to the translational field strength
through
\be
T^{\rho}{}_{\mu\nu} = h_a{}^\rho \, T^a{}_{\mu \nu}.
\ee
The Weitzenb\"ock connection can be decomposed as
\begin{equation}
\Gamma^{\rho}{}_{\mu\nu} = \Gammabol^{\rho}{}_{\mu\nu} 
+ K^{\rho}{}_{\mu\nu},
\label{rela}
\end{equation}
where $\Gammabol^{\rho}{}_{\mu\nu}$ is the Christoffel connection of the metric
$g_{\mu\nu}$, and
\begin{equation}
K^{\rho}{}_{\mu \nu} = \frac{1}{2} \left( 
T_{\mu}{}^{\rho}{}_{\nu} + T_{\nu}{}^{\rho}{}_{\mu} 
- T^{\rho}{}_{\mu \nu} \right)
\label{contorsion}
\end{equation}
is the contortion tensor. It is important to remark that we are considering
curvature and torsion as properties of a connection, not of spacetime
\cite{livro}. Notice, for example, that the Christoffel and the Weitzenb\"ock
connections are defined on the very same spacetime metric manifold.

The Lagrangian of the teleparallel equivalent of general relativity is
\begin{equation}
{\mathcal L} = \frac{c^{4} h}{16\pi G} \, S^{\rho\mu\nu}\,T_{\rho\mu\nu} +
{\mathcal L}_M,
\label{gala}
\end{equation}
where $h = {\rm det}(h^{a}{}_{\mu})$, ${\mathcal L}_M$ is the Lagrangian of a
source field, and
\begin{equation}
S^{\rho\mu\nu} = - S^{\rho\nu\mu} = \frac{1}{2} 
\left[ K^{\mu\nu\rho} - g^{\rho\nu}\,T^{\sigma\mu}{}_{\sigma} 
+ g^{\rho\mu}\,T^{\sigma\nu}{}_{\sigma} \right]
\label{S}
\end{equation}
is a tensor written in terms of the Weitzenb\"ock connection only. Performing a
variation with respect to the gauge potential, we find the teleparallel version
of the gravitational field equation \cite{sp1}
\begin{equation}
\partial_\sigma(h S_\lambda{}^{\rho \sigma}) -
\frac{4 \pi G}{c^4} \, (h t_\lambda{}^\rho) =
\frac{4 \pi G}{c^4} \, (h {\mathcal T}_\lambda{}^\rho),
\label{eqs1}
\end{equation}
where
\begin{equation}
h \, t_\lambda{}^\rho = \frac{c^4 h}{4 \pi G} \, S_{\mu}{}^{\rho \nu}
\,\Gamma^\mu{}_{\nu\lambda} - \delta_\lambda{}^\rho \, {\mathcal L}_G
\label{emt}
\end{equation}
is the energy-momentum pseudotensor of the gravitational field, and
${\mathcal T}_\lambda{}^\rho$ is the symmetric energy-momentum tensor of the
source field. It is given by ${\mathcal T}_\lambda{}^\rho = {\mathcal
T}_a{}^\rho \, h^a{}_\lambda$, with
\be
{\mathcal T}_a{}^\rho = - \frac{1}{h}
\frac{\delta {\mathcal L}_M}{\delta B^a{}_\rho} \equiv - \frac{1}{h}
\frac{\delta {\mathcal L}_M}{\delta h^a{}_\rho}.
\ee
It can be verified that, when ${\mathcal L}_M$ is written with the appropriate
spin connection \cite{tsc}, ${\mathcal T}_\lambda{}^\rho$ is in fact symmetric.
Teleparallel gravity is known to be equivalent to general relativity. In fact,
up to a divergence, when rewritten in terms of the Christoffel connection, the
Lagrangian (\ref{gala}) becomes the Einstein-Hilbert Lagrangian of general
relativity. Accordingly, the teleparallel field equation (\ref{eqs1}) is found
to coincide with Einstein's equation.

%%%%%%%%%%%%%%%%%%%%%%%%%%%%%%%%%%%%%%%%%%%%%%%%%
\section{Equation of motion: the Newtonian limit}

The equation of motion of a spinless particle of mass $m$ in teleparallel
gravity is given by the gravitational Lorentz force equation \cite{wep}
\be
h^a{}_\mu \; \frac{d u_a}{d s} =
T^b{}_{\mu \rho} \; u_b \, u^\rho,
\label{eqmot2}
\ee
where $u^a$ is the particle four-velocity seen from the tetrad frame,
necessarily anholonomic when expressed in terms of the {\it spacetime} line
element $ds = (g_{\mu \nu} dx^\mu dx^\nu)^{1/2}$. We see from this equation that
the teleparallel field strength $T^a{}_{\mu \rho}$ plays, for the 
gravitational force, the role played by the electromagnetic field strength
$F_{\mu \rho}$ in the Lorentz force. Using the relation $u_a = h_a{}^\mu \,
u_\mu$, the force equation (\ref{eqmot2}) can be rewritten in the form
\be
\frac{d u_\mu}{ds} - \Gamma^\lambda{}_{\mu \rho} \, u_\lambda \, u^\rho =
T^\lambda{}_{\mu \rho} \, u_\lambda \, u^\rho.
\label{glfe}
\ee
We notice once more that, by using the identity $T_{\lambda \mu \rho} \;
u^\lambda \, u^\rho = - K_{\lambda \mu \rho} \; u^\lambda \, u^\rho$, as well as
the relation (\ref{rela}), this equation is seen to coincide with the geodesic
equation of general relativity. On the other hand, by using Eq. (\ref{tor}), it
becomes
\be
\frac{d u_\mu}{ds} - \Gamma^\lambda{}_{\rho \mu} \, u_\lambda \, u^\rho = 0.
\label{nonco1}
\ee
As $\Gamma^\lambda{}_{\rho \mu}$ is not symmetric in the last two indices, the
left-hand side is not the covariant derivative of $u_\mu$, and consequently Eq.
(\ref{nonco1}) does not represent a geodesic line of the underlying
Weitzenb\"ock spacetime.

Let us consider now the Newtonian limit, which is obtained by assuming that the
gravitational field is stationary and weak. This means respectively that the
time derivative of $B^a{}_\mu$ vanishes, and that $|B^a{}_\mu| \ll 1$.
Accordingly, all particles will move with a sufficient small velocity so
that\footnote{We use $i, j, k, \dots = 1, 2, 3$ to denote the space components
of the holonomic (spacetime) indices.} $u^i$ can be neglected in relation
to
$u^0$. In this case, the equation of motion (\ref{nonco1}) can be written as
\be
\frac{d^2 x_\mu}{ds^2} - \Gamma_{0 0 \mu} \, c^2 \, \left(\frac{dt}{ds}
\right)^2 = 0.
\label{nonco2}
\ee
Now, taking into account that the field is stationary, up to first order in
$B^a{}_\mu$, we obtain from Eq. (\ref{carco}) that
\be
\Gamma_{0 0 \mu} = \partial_\mu B_{00}.
\ee
The equation of motion (\ref{nonco2}) is then equivalent to the two equations
\be
\frac{d^2 \vec{x}}{ds^2} = - c^2 \, \vec{\nabla} B_{00} \left(\frac{dt}{ds}
\right)^2,
\ee
and
\be
\frac{d^2 t}{ds^2} = 0,
\ee
where $x_0=x^0=ct$, and the components of $\vec{x}$ are given by $x^i = - x_i$.
The solution of the second equation is that $dt/ds$ equals a constant. Dividing
by
$(dt/ds)^2$ puts  the first equation in the form
\be
\frac{d^2 \vec{x}}{dt^2} = - c^2 \, \vec{\nabla} B_{00}.
\label{nonco3}
\ee
The corresponding Newtonian result is
\be
\frac{d^2 \vec{x}}{dt^2} = - \vec{\nabla} \phi,
\label{newton}
\ee
where $\phi$ is the gravitational potential. Comparing (\ref{nonco3}) with
(\ref{newton}) we see that
\be
c^2 \, B_{00} = \phi.
\label{bfi}
\ee

%%%%%%%%%%%%%%%%%%%%%%%%%%%%%%%%%%%%%%%%%%%
\section{Global formulation of gravitation}

As is well known, in addition to the usual {\em differential} formalism,
electromagnetism presents also a {\em global} formulation in terms of a
nonintegrable phase factor \cite{wy}. According to this approach,
electromagnetism can be considered as the gauge invariant action of a
nonintegrable (path-dependent) phase factor. For a particle with electric charge
$e$ traveling from an initial point ${\sf P}$ to a final point ${\sf Q}$, the
phase factor is given by
\be
\Phi_e({\sf P}|{\sf Q}) = \exp \left[\frac{i e}{\hbar c} \int_{\sf P}^{\sf Q}
A_\mu \, dx^\mu \right],
\ee
where $A_\mu$ is the electromagnetic gauge potential. In the classical
(non-quantum) limit, the action of this nonintegrable phase factor on a particle
wave-function yields the same results as those obtained from the Lorentz force
equation
\be
\frac{d u^a}{ds} = \frac{e}{m c^2} \, F^a{}_b \, u^b.
\ee
In this sense, the phase-factor approach can be considered as the {\em quantum}
generalization of the {\em classical} Lorentz force equation. It is actually
more general, as it can be used both on simply-connected and on
multiply-connected domains. Its use is mandatory, for example, to describe the
Bohm-Aharonov effect, a quantum phenomenon taking place in a multiply-connected
space.

Now, in the teleparallel approach to gravitation, the fundamental field
describing gravitation is the translational gauge potential $B^a{}_\mu$. Like
$A_\mu$, it is an Abelian gauge potential. Thus, in analogy with
electromagnetism, $B^a{}_\mu$ can be used to construct a global formulation for
gravitation. To start with, let us notice that the electromagnetic phase factor
$\Phi_e({\sf P}|{\sf Q})$ is of the form
\be
\Phi_e({\sf P}|{\sf Q}) = \exp \left[\frac{i}{\hbar} \, S_e \right],
\ee
where $S_e$ is the action integral describing the interaction of the charged
particle with the electromagnetic field. Now, in teleparallel gravity, the
action integral describing the interaction of a particle of mass $m$ with
gravitation is given by \cite{wep}
\be
S_g = \int_{\sf P}^{\sf Q} m \, c \, B^a{}_\mu \, u_a \, dx^\mu.
\ee
Therefore, the corresponding gravitational nonintegrable phase factor turns out
to be
\be
\Phi_g({\sf P}|{\sf Q}) = \exp \left[\frac{i m c}{\hbar} \int_{\sf P}^{\sf Q}
B^a{}_\mu \, u_a \, dx^\mu \right].
\label{npf}
\ee
Similarly to the electromagnetic phase factor, it represents the {\em quantum}
mechanical law that replaces the {\em classical} gravitational Lorentz force
equation (\ref{eqmot2}).

%%%%%%%%%%%%%%%%%%%%%%%%%%%%
\section{The COW experiment}

As a first application of the gravitational nonintegrable phase factor
(\ref{npf}), we consider the COW experiment \cite{cow}. It consists in using a
neutron interferometer to observe the quantum mechanical phase shift of neutrons
caused by their interaction with Earth's gravitational field, which is usually
assumed to be Newtonian. Now, as we have already seen, a Newtonian gravitational
field is characterized by the condition that only $B^0{}_0 \neq 0$. Furthermore,
as the experience is performed with thermal neutrons, it is possible to use the
small velocity approximation. In this case, the gravitational phase factor
(\ref{npf}) becomes
\be
\Phi_g({\sf P}|{\sf Q}) = \exp \left[\frac{i m}{\hbar} \int_{\sf P}^{\sf Q} c^2
\, B_{00} \, dt \right],
\label{npf2}
\ee
where we have used that $u^0 = \gamma \simeq 1$ for the thermal neutrons. In
the Newtonian approximation we can use the identification (\ref{bfi}), where
\be
\phi = g \, z
\ee
is the (homogeneous) Earth Newtonian potential. In this expression, $g$ is the
grav\-itational acceleration, assumed not to change significantly in the region
of the experience, and $z$ is the distance from Earth taken from some reference
point. Consequently, the phase factor can be rewritten in the form
\be
\Phi_g({\sf P}|{\sf Q}) = \exp \left[\frac{i m g}{\hbar} \int_{\sf P}^{\sf Q}
z(t) \, dt \right] \equiv \exp i \varphi.
\label{npf3}
\ee

%%%%%%%%%%%%%%%%%%%
\begin{figure}[ht]
\begin{center}
\includegraphics[height=6cm,width=6.5cm]{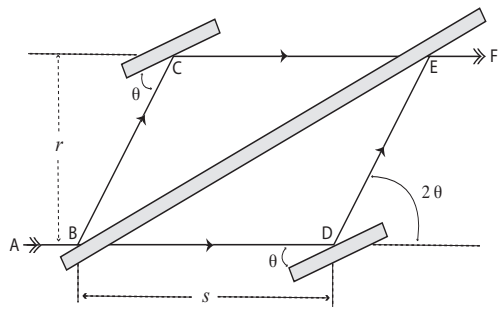}
\end{center}
\vspace{-40pt}
\caption{Schematic illustration of the COW neutron interferometer.}
\label{fig1}
\end{figure}
%%%%%%%%%%%%%%%%%%%

Let us now compute the phase $\varphi$ through the two trajectories of
Fig.~\ref{fig1}. First, we consider the trajectory {\sf BDE}. Assuming that the
segment {\sf BD} is at $z=0$, we obtain
\be
\varphi_{\sf BDE} = \frac{m g}{\hbar} \int_{\sf D}^{\sf E} z(t) \, dt.
\ee
For the trajectory {\sf BCE}, we have
\be
\varphi_{\sf BCE} = \frac{m g}{\hbar} \int_{\sf B}^{\sf C} z(t) \, dt + \frac{m
g r}{\hbar} \int_{\sf C}^{\sf E} dt.
\ee
As the phase contribution along the segments {\sf DE} and {\sf BC} are equal,
we get
\be
\Delta \varphi \equiv \varphi_{\sf BCE} - \varphi_{\sf BDE} =
\frac{m g r}{\hbar} \int_{\sf C}^{\sf E} dt.
\ee
Now, since the neutron velocity is constant along the segment {\sf CE}, we have
that
\be
\int_{\sf C}^{\sf E} dt \equiv \frac{s}{v} = \frac{s m \lambda}{h},
\ee
where $s$ is the length of the segment {\sf CE}, and $\lambda = h/(m v)$ is the
de Broglie wavelength associated with the neutron. We thus obtain
\be
\Delta \varphi =  s \, \frac{2 \pi g r \lambda m^2}{h^2},
\ee
which is exactly the gravitationally induced phase difference predicted for the
COW experience \cite{cow}. It is important to remark that, in the above
calculation, we have used the weak equivalence principle, according to which the
gravitational ($m_g$) and the inertial ($m_i$) masses are assumed to be equal.
If they were supposed to be different, the phase shift would be given by
\be
\Delta \varphi = s \, \frac{2 \pi g r \lambda m_g m_i}{h^2}.
\ee

%%%%%%%%%%%%%%%%%%%%%%%%%%%%%%%%%%%%%%%%%%%%
\section{Gravitational Aharonov-Bohm effect}

As a second application we use  the phase factor (\ref{npf}) to study the
gravitational analog of the Aharonov-Bohm effect \cite{gabere}. The usual
(electromagnetic) Aharonov-Bohm effect consists in a shift, by a constant
amount, of the electron interferometry wave pattern, in a region where there is
no magnetic field, but there is a nontrivial gauge potential $A_i$. Analogously,
the gravitational Aharonov-Bohm effect will consist in a similar shift of the
same wave pattern, but produced by the presence of a gravitational gauge
potential $B_{0 i}$. Phenomenologically, this kind of effect might be present
near a massive rapidly rotating source, like a neutron star, for example. Of
course, differently from an ideal apparatus, in a real situation the
gravitational field cannot be eliminated, and consequently the gravitational
Aharonov-Bohm effect should be added to the other effects also causing a phase
change.

\begin{figure}[ht]
\begin{center}
\includegraphics[height=5.5cm,width=9.5cm]{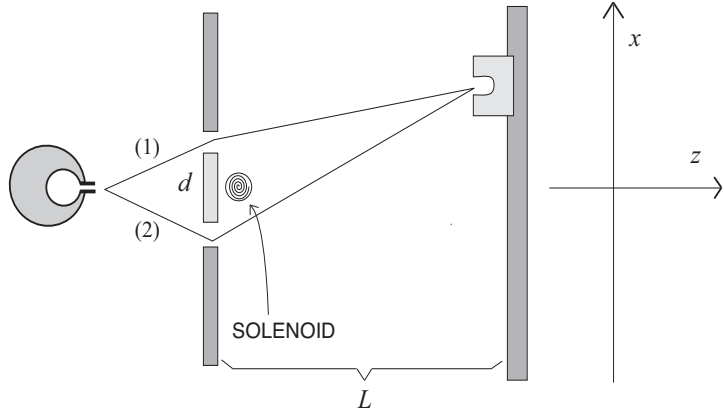}
\end{center}
\caption{Schematic illustration of the Aharonov-Bohm electron interferometer.}
\label{fig2}
\end{figure}

Let us consider first the case in which there is no external field at all. If
the electrons are emitted with a characteristic momentum $p$, then its
wavefunction has the de Broglie wavelength $\lambda = h/p$. Denoting by $L$ the
distance between slit and screen (see Fig.~\ref{fig2}), and by $d$ the distance
between the two holes, when the conditions $L \gg \lambda$, $L \gg x$ and $L \gg
d$ are satisfied, the phase difference at a distance $x$ from the central point
of the screen is given by
\be
\delta^0 \varphi(x) = \frac{2 \pi x d}{L \lambda}.
\label{wpat}
\ee
This expression defines the wave pattern on the screen.

%%%%%%%%%%%%%%%%%%%
%%%%%%%%%%%%%%%%%%%

We consider now the case in which a kind of infinite ``gravitational solenoid''
produces a purely static gravitomagnetic field flux concentrated in its
interior. In the ideal situation, the gravitational field outside the solenoid
vanishes completely, but there is a nontrivial gauge potential $B_{0 i}$. When
we let the electrons to move outside the solenoid, phase factors corresponding
to paths lying on one side of the solenoid will interfere with phase factors
corresponding to paths lying on the other side, which will produce an additional
phase shift at the screen. Let us then calculate this additional phase shift.
The gravitational phase factor (\ref{npf}) for the physical situation described
above is
\be
\Phi_g({\sf P}|{\sf Q}) = \exp \left[- \frac{i m c}{\hbar} \int_{\sf P}^{\sf Q}
u^0 \vec{B}_0 \cdot d\vec{r} \right],
\ee
where $\vec{B}_0$ is the vector with components $B_0{}^i = - B_{0i}$. Since
$u^0 = \gamma \equiv [1 - ({v^2}/{c^2}) ]^{-1/2}$, and considering that the
electron velocity $v$ is constant, we can write
\be
\Phi_g({\sf P}|{\sf Q}) = \exp \left[- \frac{i \gamma m c}{\hbar} \int_{\sf
P}^{\sf Q} \vec{B}_0 \cdot d\vec{r} \right].
\ee
Now, denoting by $\varphi_1$ the phase corresponding to a path lying on one
side of the solenoid, and by $\varphi_2$ the phase corresponding to a path lying
on the other side, the phase difference at the screen will be
\be
\delta \varphi \equiv \varphi_2 -\varphi_1 =
\frac{\gamma m c}{\hbar} \oint \vec{B}_0 \cdot d\vec{r} =
\frac{{\mathcal E} \, \Omega}{\hbar \, c},
\label{gabe}
\ee
where ${\mathcal E} = \gamma m c^2$ is the electron kinetic energy, and
\be
\Omega = \oint \vec{B}_0 \cdot d\vec{r} = \oint (\vec{\nabla} \times \vec{B}_0)
\cdot d\vec{\sigma} \equiv \oint \vec{H} \cdot d\vec{\sigma}
\ee
is the $\vec{H}$ flux inside the solenoid. In components, the gravitational field
$\vec{H}$ is written as
\be
H^i = \frac{1}{2} \epsilon^{ijk} \; (\partial_j B_{0 k} - \partial_k B_{0 j})
= \frac{1}{2} \epsilon^{ijk} \; T_{0 j k},
\label{axtor}
\ee
with $F_{0 j k}$ the gravitational field strength defined in (\ref{gfs}).
Remembering that the axial torsion is defined by ${\mathcal A}^\mu = ({1}/{6})
\epsilon^{\mu \nu \rho \sigma} \, T_{\nu \rho \sigma}$, it is then an easy task
to verify that $H^i = {\mathcal A}^i$, with
\be
{\mathcal A}^i = - \frac{1}{2} \epsilon^{0 i j k} \, T_{0 j k}
\ee
the space components of the axial torsion. This shows that $H^i$ coincides with
the axial torsion, which in turn represents the gravitomagnetic component of the
gravitational field
\cite{kerr}.

Expression (\ref{gabe}) gives the phase difference produced by the interaction
of the particle's kinetic energy with a gauge potential, which gives rise to the
gravitational Aharonov-Bohm effect. As this phase difference depends on the
energy, it applies equally to massive and massless particles. It is worthy
mentioning that, for the case of massive particles, if the inertial and
gravitational masses were supposed to be different, the phase shift would assume
the form
\be
\delta \varphi = \frac{m_g}{m_i} \,
\frac{{\mathcal E} \, \Omega}{\hbar \, c}.
\ee
It is also worthy mentioning that, whereas for massive particles it is a
genuine quantum effect, for massless particles, due to the their intrinsic wave
character, it can be considered as a classical effect. In fact, for
${\mathcal E}=h \omega$, Eq.~(\ref{gabe}) becomes
\be
\delta \varphi =
\frac{\omega \, \Omega}{c},
\label{gabeb}
\ee
and we see that, in this case, the phase difference does not depend on the
Planck's constant.

In contrast with the electromagnetic Aharonov-Bohm effect, the phase difference
in the gravitational case depends on the particle kinetic energy, which in turn
depends on the particle's mass and velocity. Like the electromagnetic case,
however, the phase difference is independent of the position $x$ on the screen,
and consequently the wave pattern defined by (\ref{wpat}) will be whole shifted
by a constant amount. It is also important to mention that the requirement of
invariance of the gravitational phase factor under the translational gauge
transformation (\ref{btrans}) implies that
\be
\Omega {\mathcal E} = n h c,
\ee
with $n$ an integer number. Differently from the electromagnetic case, where it
is pos\-sible to define a {\em quantum} of magnetic flux, we see from the above
equation that in the gravitational case it is not possible to define a
particle-independent {\em quantum} of gravitomagnetic flux \cite{harris}.

%%%%%%%%%%%%%%%%%%%%%%%%%%%%%%%%%%%%%%%%%%%%%
\section{Quantum versus classical approaches}
\label{qcg}

\begin{figure}[ht]
\begin{center}
\includegraphics[height=5.5cm,width=9cm]{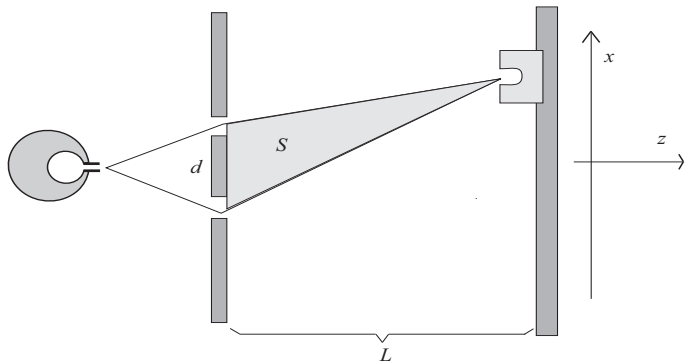}
\end{center}
\vspace{-20pt}
\caption{Schematic illustration of the electron interference experiment in a
gravitomagnetic field. The only contribution to the phase-shift comes from the
flux through the surface $S$ delimited by the two trajectories.}
\label{fig3}
\end{figure}

We proceed now to show that, in the classical limit, the nonintegrable phase
factor approach reduces to the usual approach provided by the gravitational
Lorentz force equation. In  electromagnetism, the standard argument is
well-known: the phase turning up in the quantum case is exactly the classical
action, which leads to the Lorentz force. We intend, however, to illustrate the
result directly, and for that we consider again the electron interferometry slit
experiment, this time with a homogeneous static gravitomagnetic field $\vec{H}$
permeating the whole region between the slit and the screen (see
Fig.~\ref{fig3}). This field is supposed to point in the negative $y$-direction,
and will produce a phase shift which is to be added to the phase (\ref{wpat})
extant in the absence of gravitomagnetic field. This shift, according to
Eq.~(\ref{gabe}), is given by ${\mathcal E} \Omega/\hbar c$, with $\Omega$ the
flux through the surface $S$ circumscribed by the two trajectories.
%%%%%%%%%%%%%%%%%%%
%%%%%%%%%%%%%%%%%%%
It is easily seen that $S = L d/2$ for any value of $x$. The flux is
consequently
\be
\Omega = - H_y L d/2,
\ee
where $\vec{H} = - H_y \, \hat{e}_y$, with $\hat{e}_y$ a unity vector in the
$y$ direction. Therefore, the total phase difference will be
\be
\delta \varphi = \frac{2 \pi x d}{L \lambda} - \frac{{\mathcal E} H_y L d}{2
\hbar c}.
\ee
This is the result yielded by the phase-factor approach.

In the classical limit, the slit experiment can be interpreted in the following
way. The electrons traveling through the gravitomagnetic field have their
movement direction changed. This means that they experiment a force in the
$x$-direction. For small $x$, we can approximately write the electrons velocity
as $v \simeq v_z$. In this case, they will be transversally accelerated by the
gravitomagnetic field during the time interval
\be
\Delta t = \frac{L}{v_z}. 
\ee
This transversal $x$-acceleration is given by
\be
a_x = \frac{2 x}{(\Delta t)^2}.
\ee
Since the attained acceleration is constant, we can choose a specific point to
calculate it. Let us then consider the point of maximum intensity on the screen,
which is determined by the condition
$\delta \varphi = 0$. This yields
\be
x = \frac{H_y \, \lambda \, L^2 \, {\mathcal E}}{2\, h \, c}.
\label{ex}
\ee
The acceleration is then found to be
\be
a_x = \, \frac{H_y \, \lambda \, {\mathcal E} \, v_z^2}{h \, c}.
\ee
From the classical point of view, therefore, we can say that the electrons
experience a force in the $x$-direction given by
\be
{\mathcal F}_x \equiv \gamma \, m \, a_x =
\frac{{\mathcal E} \, v_z \, H_y}{c} \, \frac{\lambda \, p}{h}, 
\ee
with $p = \gamma m v_z$ the electron momentum. Using now the de Broglie
relation $\lambda = h/p$, it is possible to eliminate the Planck constant. We
then get the classical result
\be
{\mathcal F}_x = \frac{\mathcal E}{c} \, v_z H_y =
 \frac{{\mathcal E}}{c} \; (\vec{v} \times \vec{H} )_x,
\ee
which gives rise to the equation of motion
\be
\frac{D p_x}{dt} =  \frac{{\mathcal E}}{c} \; (\vec{v} \times \vec{H} )_x,
\ee
where $D/dt$ represents a time covariant derivative in the Weitzenb\"ock
connection. This is exactly the $x$-component of the gravitational Lorentz force
equation (\ref{eqmot2}) (see Appendix A). From this equation we see clearly
that, when the inertial and gravitational masses coincide, the resulting
trajectory does not depend on the mass of the particle, which is in accordance
with the universality property of the gravitational interaction. It is important
to remark that, even if the inertial and gravitational masses are not supposed
to coincide, the teleparallel approach still yields a consistent description of
the gravitational interaction
\cite{wep}

%%%%%%%%%%%%%%%%%%%%%%%
\section{Final remarks}

Teleparallel gravity is a gauge theory for the translation group. Its
fundamental field is, accordingly, a gauge potential $B^a{}_\mu$  with values in
the Lie algebra of the translation group. In this formulation, gravitation
becomes quite analogous to electromagnetism. Based on that analogy  and relying
on the phase-factor approach to Maxwell's theory, a teleparallel nonintegrable
phase-factor approach to gravitation has been developed, which represents the
quantum mechanical version of the classical gravitational Lorentz force.

As a first application  we have considered the COW experiment. By taking the
Newtonian limit, we have shown that the formalism yields the correct quantum
phase-shift induced on the neutrons by their interaction with Earth's
gravitational field. As this phase shift is produced by the coupling of the
neutron mass with the component $B_{00}$ of the translational gauge potential,
it can be considered as a gravitoelectric Aharonov-Bohm effect \cite{pt}. As a
second application  we have obtained the quantum phase-shift produced by the
coupling of the particle's kinetic energy with the components $B_{0i}$ of the
gauge potential, which corresponds to the usual (gravitomagnetic) analog of the
Aharonov-Bohm effect.

Finally, by considering a simple slit experiment in which the particles are
allowed to pass through a region with a homogeneous gravitomagnetic field, we
have shown that, in the classical limit, the (quantum) nonintegrable
phase-factor approach coincides with the usual (classical) approach based on the
gravitational Lorentz force equation. This means that, as far as only classical
trajectories are concerned, both approaches give the same result. In addition,
provided the gravitational and inertial masses are supposed to be equal,
gravitation becomes universal in the classical limit in the sense that the
resulting trajectories do not depend on the masses of the particles. 

At the quantum level, however, there are deep conceptual changes with respect
to classical gravity. The phase of the  particle wavefunction acquires a
fundamental status and depends on the particle mass (COW effect, obtained in the
non-relativistic limit) or relativistic kinetic energy (gravitational
Aharonov-Bohm effect). At the quantum level, therefore, gravitation is no more
universal \cite{nonuni}, although in the specific case of the non-relativistic
COW experiment, by introducing a kind of quantum equivalence principle, it can
be made independent of the mass when written in an appropriate way
\cite{lammerzahl}. Concerning this point, it should be remarked that, since
teleparallel gravity is able to describe gravitation independently of the
validity or not of the equivalence principle \cite{wep}, it will not require a
quantum version of this principle to deal with gravitationally induced quantum
effects. We can thus conclude that teleparallel gravity seems to provide a much
more appropriate and consistent approach to study these effects.

%%%%%%%%%%%%%%%%%%%%%%%%%%
\ack
The authors would like to thank FAPESP-Brazil, CNPq-Brazil, and CAPES-Brazil for
financial support.

%%%%%%%%%
\appendix
\section{}

Let us take the gravitational Lorentz force equation (\ref{eqmot2}) written in
the form
\be
\frac{Du_\mu}{d s} \equiv \frac{d u_\mu}{d s} -
\Gamma^\lambda{}_{\mu \rho} \, u_\lambda \, u^\rho = T_{\lambda \mu \rho} \,
u^\lambda \, u^\rho.
\ee
As already mentioned, it is equivalent to the geodesic equation of general
relativity. Its left-hand side is the Weitzenb\"ock covariant derivative of the
four-velocity along the trajectory, and the right-hand side represents the
gravitational force. In the specific case considered in section~\ref{qcg}, where
a gravitomagnetic field is present, only the components $T_{0ij}$ of torsion
are nonvanishing. Consequently, the above force equation reduces to
\be
\frac{Du_i}{d s} = T_{0 i j} \, u^0 \, u^j.
\ee
Using the relations $u^0 = \gamma$ and $u^j = (\gamma v^j)/c$, as well as the
fact that the gravitomagnetic field does not change the absolute value of the
particle velocity, and consequently $d \gamma/ds = 0$, we obtain
\be
\frac{Dv_i}{d s} = \gamma \, c \, T_{0 i j} \, v^j.
\ee
Now, from Eq. (\ref{axtor}) we see that $T_{0 i j} = \epsilon_{ijk} \, H^k$.
Therefore, by using that $ds = (c/\gamma) dt$, we get
\be
\frac{Dv_i}{d t} = c \, \epsilon_{ijk} \, v^j \, H^k =
c \, (\vec{v} \times \vec{H})_i.
\ee
The corresponding equation for the momentum components $p^i = \gamma m v^i$ is
\be
\frac{Dp^i}{d t} = \frac{\mathcal E}{c} \, (\vec{v} \times \vec{H})^i.
\ee
The force appearing in the right-hand side is quite similar to the
electromagnetic Lorentz force, with the kinetic energy replacing the electric
charge, and the gravitomagnetic vector field $\vec{H}$ replacing the usual
magnetic field.
 
%%%%%%%%%%%%%%%%%%%%%%%%%%%

\end{document}